\documentclass{elsarticle}
\usepackage{lineno}
\usepackage{epsfig}
\usepackage{multirow}
\usepackage{upgreek}
\usepackage{subcaption}
\usepackage{adjustbox}
\usepackage[utf8]{inputenc}
\usepackage[T1]{fontenc}

\begin{document}

\newcommand{\mum}{\ensuremath{\,\upmu\mathrm{m}}}
\newcommand{\vgl}{\ensuremath{V_\mathrm{gl}}}
\newcommand{\vfd}{\ensuremath{V_\mathrm{FD}}}
\newcommand{\vbias}{\ensuremath{V_\mathrm{bias}}}
\newcommand{\gfixed}{\ensuremath{G_{100\,\mathrm{V}}}}
\newcommand{\strontium}{\ensuremath{^{90}\mathrm{Sr}}}
\newcommand{\neqcm}{\ensuremath{\,\mathrm{n}_\mathrm{eq}\,\mathrm{cm}^{-2}}}
\newcommand{\up}[1]{\ensuremath{\,\mathrm{#1}}}

\begin{frontmatter}
\title {TCT-based monitoring of LGAD radiation hardness for ATLAS-HGTD production}



\author[ljubljana]{I.~Velkovska}


\author[rabat]{A.~Aboulhorma}
\author[rabat]{M.~Ait~Tamlihat}
\author[tdli,sjtu]{H.M.~Alfanda}
\author[jinr]{O.~Atanova}
\author[jinr]{N.~Atanov}
\author[mascir]{I.~Azzouzi}
\author[ihep]{J.~Barreiro~Guimar\~{a}es~da~Costa}
\author[lpnhe]{T.~Beau}
\author[casablanca]{D.~Benchekroun}
\author[casablanca]{F.~Bendebba}
\author[cern]{G.~Bergamin}
\author[benguerir]{Y.~Bimgdi}
\author[ijclab]{A.~Blot}
\author[jinr]{A.~Boikov}
\author[ijclab]{J.~Bonis}
\author[lpca]{D.~Boumediene}
\author[lip]{C.~Brito}
\author[mainz]{A.S.~Brogna}
\author[cern]{A.M.~Burger}
\author[ijclab]{L.~Cadamuro}
\author[nju]{Y.~Cai}
\author[lpca]{N.~Cartalade}
\author[ifae]{R.~Casanova~Mohr}
\author[rabat]{R.~Cherkaoui~El~Moursli}
\author[nju]{Y.~Che}
\author[sjtu]{X.~Chen}
\author[nijmegen]{E.Y.S.~Chow}
\author[lpca]{L.D.~Corpe}
\author[lpca]{C.G.~Crozatier}
\author[lpca]{L.~D'Eramo}
\author[taipei]{S.~Dahbi}
\author[cern]{D.~Dannheim}
\author[lpnhe]{G.~Daubard}
\author[jinr]{Y.~Davydov}
\author[ljubljana]{J.~Debevc}
\author[cea]{Y.~Degerli}
\author[cea]{E.~Delagnes}
\author[cea]{F.~Deliot}
\author[lpnhe]{M.~Dhellot}
\author[omega]{P.~Dinaucourt}
\author[cern]{G.~Di~Gregorio}
\author[lip]{P.J.~Dos~Santos~De~Assis}
\author[ihep]{C.~Duan}
\author[ijclab]{O.~Duarte}
\author[omega]{F.~Dulucq}
\author[mainz]{J.~Ehrecke}
\author[hefei]{Y.~El~Ghazali}
\author[casablanca]{A.~El~Moussaouy}
\author[ijclab]{A.~Falou}
\author[ihep]{L.~Fan}
\author[ihep]{Y.~Fan}
\author[ihep]{Z.~Fan}
\author[taipei]{K.~Farman}
\author[rabat]{F.~Fassi}
\author[ihep]{Y.~Feng}
\author[lip]{M.~Ferreira}
\author[nijmegen]{F.~Filthaut}
\author[mainz]{F.~Fischer}
\author[ifae]{P.~Fust\'e}
\author[ihep]{J.~Fu}
\author[ifae]{J.~Garci\'ia Rodriquez}
\author[lip]{G.~Gaspar~De~Andrade}
\author[ifae]{V.~Gautam}
\author[nju]{Z.~Ge}
\author[lip]{R.~Gon\c{c}alo}
\author[kenitra]{M.~Gouighri}
\author[ifae]{S.~Grinstein}
\author[jinr]{K.~Gritsay}
\author[cea]{F.~Guilloux}
\author[cern]{S.~Guindon}
\author[lpca]{A.~Haddad}
\author[ijclab]{S.E.D.~Hammoud}
\author[nju]{L.~Han}
\author[cern]{A.M.~Henriques~Correia}
\author[kenitra]{M.~Hidaoui}
\author[ljubljana]{B.~Hiti\corref{cor1}}\cortext[cor1]{Corresponding author}
\author[mainz]{J.~Hofner}
\author[taipei]{S.~Hou}
\author[ihep,ucas]{X.~Huang}
\author[ihep]{Y.~Huang}
\author[sdu]{K.~Hu}
\author[lpca]{C.~Insa}
\author[ijclab]{J.~Jeglot}
\author[ihep,ucas]{X.~Jia}
\author[ljubljana]{G.~Kramberger}
\author[usp]{M.~Kuriyama}
% ...

\affiliation[ifae]{organization={Institut de F\'isica d'Altes Energies (IFAE), Barcelona Institute of Science and Technology}, city={Barcelona}, country={Spain}}

\affiliation[ihep]{organization={Institute of High Energy Physics, Chinese Academy of Sciences}, city={Beijing}, country={China}}
\affiliation[ucas]{organization={University of Chinese Academy of Sciences (UCAS)}, city={Beijing}, country={China}}

\affiliation[usp]{organization={Instituto de F\'isica, Universidade de S\~ao Paulo}, city={S\~ao Paulo}, country={Brazil}}

\affiliation[cern]{organization={CERN}, city={Geneva}, country={Switzerland}}

\affiliation[nju]{organization={Department of Physics, Nanjing University}, city={Nanjing}, country={China}}

\affiliation[sdu]{organization={Institute of Frontier and Interdisciplinary Science and Key Laboratory of Particle Physics and Particle Irradiation (MOE), Shandong University}, city={Qingdao}, country={China}}

\affiliation[sjtu]{organization={State Key Laboratory of Dark Matter Physics, School of Physics and Astronomy, Shanghai Jiao Tong University, Key Laboratory for Particle Astrophysics and Cosmology (MOE), SKLPPC}, city={Shanghai}, country={China}}
\affiliation[tdli]{oragnization={State Key Laboratory of Dark Matter Physics, Tsung-Dao Lee Institute, Shanghai Jiao Tong University}, city={Shanghai}, country={China}}

\affiliation[lpca]{organization={LPC, Universit\'e Clermont Auvergne, CNRS/IN2P3}, city={Clermont-Ferrand}, country={France}}

% ...
\affiliation[hefei]{organization={Department of Modern Physics and State Key Laboratory of Particle Detection and Electronics, University of Science and Technology of China}, city={Hefei}, country={China}}

\affiliation[ijclab]{organization={IJCLab, Universit\'e Paris-Saclay, CNRS/IN2P3}, city={Orsay}, country={France}}

\affiliation[jinr]{organization={Affiliated with an international laboratory covered by a cooperation agreement with CERN}}

\affiliation[ljubljana]{organization={Department of Experimental Particle Physics, Jo\v{z}ef Stefan Institute and Department of Physics, University of Ljubljana}, city={Ljubljana}, country={Slovenia}}

\affiliation[mainz]{organization={Institut f\"{u}r Physik, Universit\"{a}t Mainz}, city={Mainz}, country={Germany}}

\affiliation[mascir]{organization={Moroccan Foundation for Advanced Science Innovation and Research (MAScIR)}, city={Rabat}, country={Morocco}}

\affiliation[casablanca]{organization={Facult\'e des Sciences Ain Chock, Universit\'e Hassan II de Casablanca}, city={Casablanca}, country={Morocco}}
\affiliation[kenitra]{organization={Facult\'{e} des Sciences, Universit\'{e} Ibn-Tofail}, city={K\'{e}nitra}, country={Morocco}}
% ...
\affiliation[rabat]{organization={Facult\'e des sciences, Universit\'e Mohammed V}, city={Rabat}, country={Morocco}}
\affiliation[benguerir]{organization={Institute of Applied Physics, Mohammed VI Polytechnic University}, city={Ben Guerir}, country={Morocco}}

\affiliation[nijmegen]{organization={Institute for Mathematics, Astrophysics and Particle Physics, Radboud University/Nikhef}, city={Nijmegen}, country={Netherlands}}

% ...
\affiliation[omega]{organization={OMEGA, Ecole Polytechnique, CNRS/IN2P3}, city={Palaiseau}, country={France}}

\affiliation[lpnhe]{organization={LPNHE, Sorbonne Universit\'e, Universit\'e Paris Cit\'e, CNRS/IN2P3}, city={Paris}, country={France}}

\affiliation[lip]{organization={Laborat\'orio de Instrumenta\c{c}\~ao e F\'isica Experimental de Part\'iculas - LIP}, city={Lisboa}, country={Portugal}}
% ...
\affiliation[cea]{organization={IRFU, CEA, Universit\'e Paris-Saclay}, city={Gif-sur-Yvette}, contry={France}}

% ...
\affiliation[taipei]{organization={Institute of Physics, Academia Sinica}, city={Taipei}, country={Taiwan}}

\begin{abstract}

Production of the High Granularity Timing Detector for the ATLAS experiment at High Luminosity LHC requires over $21\,000$ silicon sensors based on Low Gain Avalanche Diode (LGAD) technology. Their radiation hardness is monitored as a part of the production quality control. Dedicated test structures from each wafer are irradiated with neutrons and a fast and comprehensive characterization is required. We introduce a new test method based on Transient Current Technique (TCT) performed in the interface region of two LGAD devices. The measurement enables extraction of numerous sensor performance parameters, such as LGAD gain layer depletion voltage, LGAD gain dependence on bias voltage, sensor leakage current and effective interpad distance. Complementary capacitance-voltage measurements and charge collection measurements with $\strontium$ on the same samples have been performed to calibrate the TCT results in terms of charge collection and define acceptance criteria for wafer radiation hardness in the ATLAS-HGTD project.

\vskip 0.6cm
 PACS: 85.30.De; 29.40.Wk; 29.40.Gx
\begin{keyword}
ATLAS, HGTD, Low Gain Avalanche Diodes, irradiations, time resolution, charge multiplication, Transient Current Technique
\end{keyword}       
\end{abstract}
\end{frontmatter}


\section{Introduction}

The Large Hadron Collider (LHC) at CERN will undergo a High Luminosity upgrade (HL-LHC) in the years 2026--2030, which will increase its instantaneous luminosity to 5 to 7.5$\times10^{34}\up{cm}^{-2}\up{s}^{-1}$, a factor of 5 to 7.5 above the original design value \cite{hl-lhc}.
To cope with increased luminosity and resulting detector occupancy (pile-up), the ATLAS experiment introduces a new High Granularity Timing Detector (ATLAS-HGTD), which enhances track-vertex association and improves pile-up rejection based on a measurement of the time of arrival of collision products \cite{atlas-hgtd-tdr}. ATLAS-HGTD provides a track time resolution below $50\up{ps}$ and is designed to withstand a fluence of $2.5\times10^{15}\neqcm$ and a total ionizing dose of $2\up{MGy}$. The innermost layers of ATLAS-HGTD will receive radiation damage above these levels and partial replacements of these components are foreseen.

The timing resolution is limited in part by electronic jitter, which arises from signal fluctuations due to noise and results in variations in the discriminator output timing.
To mitigate the impact of jitter, ATLAS-HGTD uses silicon sensors with internal gain based on Low Gain Avalanche Diode (LGAD) technology \cite{GP2014}, which allow detector operation at a high signal-to-noise ratio in the range of 15--30. Each sensor consists of a matrix of $15\times15$ pixels with a pixel size of $1.3\up{mm}\times1.3\up{mm}$, an active thickness of $50\mum$ and a total physical thickness of $775\mum$ to ensure mechanical stability.
The ATLAS-HGTD design incorporates four sensor layers to ensure that each track passing through the detector generates at least two detected hits. The sensors will be operated at $-30\,^\circ\mathrm{C}$.
Internal gain in the sensor degrades with radiation damage due to radiation induced acceptor removal in the LGAD gain layer, leading to degraded time resolution\cite{lgad_irrad1,lgad_irrad2,lgad_irrad3}. The loss of gain will be partially compensated by increasing the reverse bias voltage. However, the maximal allowed operating bias voltage on these sensors is limited to $550\up{V}$ (corresponding to an average electric field of $11\up{V}/\upmu\mathrm{m}$), to prevent destructive single-event-burnouts, which occur at electric fields $>12\up{V}/\upmu\mathrm{m}$ as a result of electrical discharges in the sensor during rare events with massive charge deposition \cite{seb1,seb2,seb3}.
The sensor will be read out by a custom chip named ALTIROC \cite{altiroc}, containing pixel cells with an analog front-end amplifier, constant fraction discriminator and time-to-digital converter. ALTIROC is designed to achieve a hit time resolution of $70\up{ps}$ for input signals corresponding to a collected charge of $4\up{fC}$ ($25\,000$ electrons), after receiving an end-of-life total ionizing doze of $2\up{MGy}$. By combining hit time information from two or more detector layers, the track time resolution will be improved to below $50\up{ps}$.
At the end-of-life fluence of $2.5\times10^{15}\neqcm$, ATLAS-HGTD sensors must hence provide a most probable charge signal of at least $4\up{fC}$ and a time resolution of $50\up{ps}$ (measured with unirradiated discrete readout electronics on a test bench) for a minimum ionizing particle (MIP). 

The sensors come in two designs, named after the respective design teams, developed independently but sharing the same specifications: IHEP-IME\footnote{IHEP: Institute of High Energy Physics, Chinese Academy of Sciences, Beijing, China} and USTC-IME\footnote{USTC: University of Science and Technology of China, Hefei, China}, both produced by IMECAS\footnote{IMECAS: Institute of Microelectronics, Chinese Academy of Sciences, Beijing, China}. The sensors are fabricated on 8-inch silicon wafers, each containing 52 sensor arrays, each with a dedicated test structure. ATLAS-HGTD will be built from $16\,064$ sensors, with additional spares included to accommodate for an expected assembly yield of $75\,\%$, bringing the overall total to over $21\,000$ sensors. This corresponds to about 1000 sensor wafers, assuming a  production yield of $40\,\%$. The large scale of the production requires a comprehensive quality control. 

One area of production quality control is sensor radiation hardness, which will be monitored in a procedure called the Irradiation Test (IT). The IT is performed on dedicated test structures from sensor wafers with a default frequency of one test per wafer. Its primary goal is verifying the compliance with the radiation hardness requirements. A secondary goal is to group samples with similar bias voltage evolution after irradiation -- in ATLAS-HGTD two sensors share a single high voltage channel and it is crucial that their breakdown voltage matches to within a few volts. 

For a reasonably fast processing, on the level of 2 hours per IT including sample handling, we developed a method based on Transient Current Technique (TCT), which will be used to extract four sensor performance parameters: LGAD gain layer depletion voltage, LGAD gain factor at a fixed bias voltage, leakage current, and width of the inactive interpad region.
The extracted parameters are compared to predefined threshold values. Passing the IT indicates that the wafer meets the radiation hardness criterion for acceptance. Devices that fail the test prompt additional investigations using TCT and charge collection measurement with $\strontium$ to determine the final decision regarding wafer acceptance or rejection. This paper presents the concept of the TCT-based IT method, procedure for sensor parameter extraction, test results with preproduction samples of ATLAS-HGTD sensors, and calibration of the method against other sensor characterization techniques (C-V and $\strontium$ radioactive source measurements).


\section{Samples and experimental technique} 

\begin{figure}[!hbt]
\centering
\begin{subfigure}[c]{0.54\textwidth}
\includegraphics[width=\columnwidth, trim=0 0 400 0, clip]{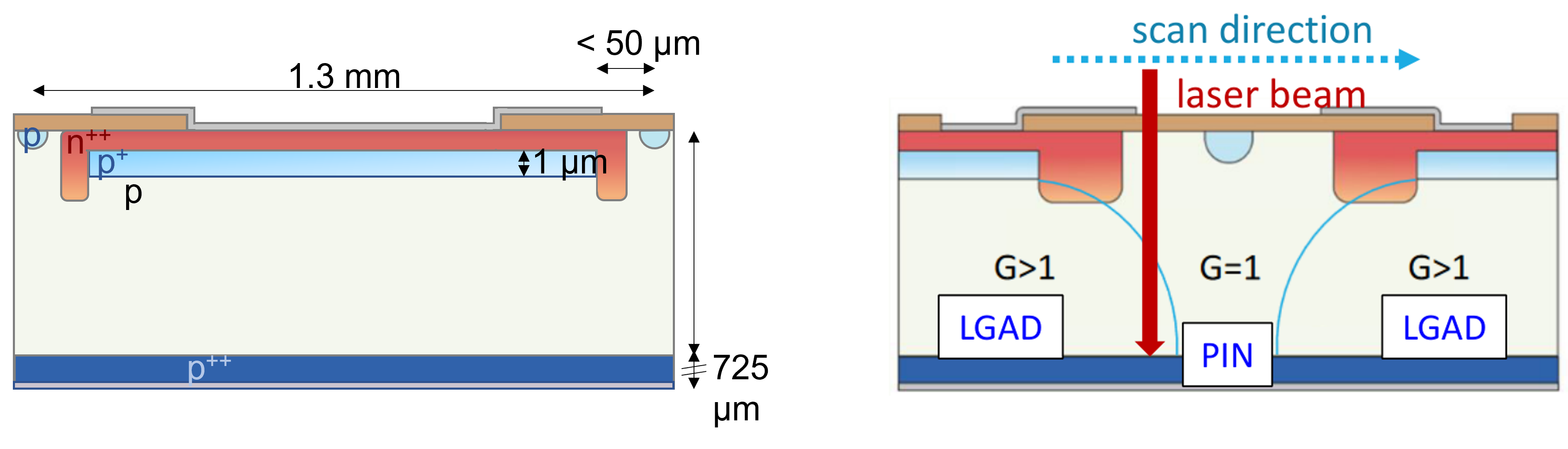}%
\caption{}
\label{fig:lgad}
\end{subfigure}
\hfill
\begin{subfigure}[c]{0.45\textwidth}
\includegraphics[width=\columnwidth, trim=470 0 0 0, clip]{plots/LGAD_both.png}%
\caption{}
\label{fig:lgad_interpad}
\end{subfigure}
\caption{
a) Schematics of an LGAD pixel designed for ATLAS-HGTD, illustrating the p$^+$ gain layer embedded between the n$^{++}$ collection electrode and the p-type silicon bulk. At the pixel edge the n$^{++}$ structure is extended to terminate the p-n junction and prevent electrical breakdown, and a p-stop structure is included for inter-pixel isolation (dimensions not to scale).
b) Concept of the TCT measurement between two LGAD pixels used in the Irradiation Test. A focused laser beam enters the structure from the top and deposits charge either within the LGADs (Gain $>1$) or in the interpad region (PIN) with no gain ($\mathrm{G}=1$).
}
\label{fig:lgad_schematics}
\end{figure}

LGAD is a type of depleted silicon sensor produced on a high resistivity p-type substrate with segmented n-type collection electrodes and a highly doped p$^+$ layer implanted below each collection electrode (Figure \ref{fig:lgad}). The p$^+$ layer has a typical thickness of $1\mum$, while the typical thickness of the p-type substrate is tens of $\mum$. The electric field in the p$^+$ layer, also called multiplication or gain layer, is sufficient to generate charge multiplication by impact ionization, while still small enough to constrain the operation to a linear regime with gain factors in the range of 10--50. 
The voltage required to fully deplete the gain layer is called $\vgl$ and is a key parameter in the characterization of LGAD performance.
On the edges of each LGAD pixel, the gain layer is interupted and an inter-pixel insulation structure is implanted. Charge carriers deposited in this volume, called \textit{interpad region}, do not undergo multiplication and yield smaller signals, which are similar to those from a standard (PIN) diode. The effective width of the interpad region in ATLAS-HGTD needs to be below $100\mum$ to ensure a reasonable geometric fill factor above $85\up{\%}$ in each of the four sensor layers. 

\begin{figure}[!hbt]
\centering
\begin{subfigure}[c]{\textwidth}
\includegraphics[width=\columnwidth]{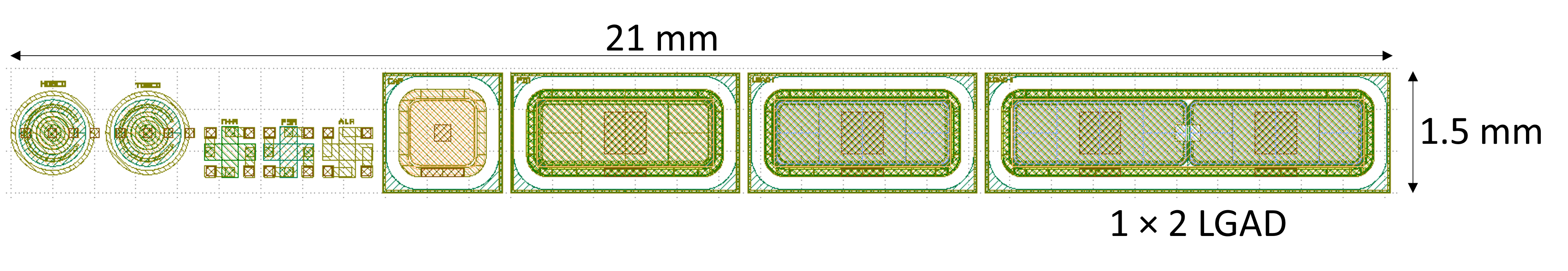}%
\caption{}
\end{subfigure}
\vspace{0.2cm}\\
\begin{subfigure}[c]{0.495\textwidth}
\includegraphics[width=\columnwidth, trim=20 12 20 15, clip]{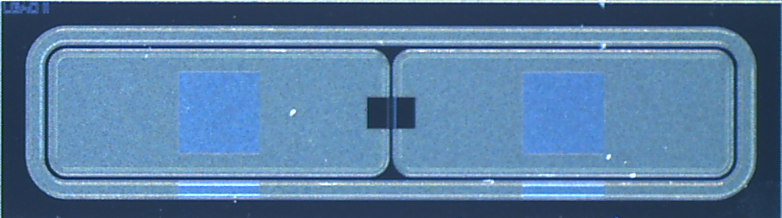}%
\caption{IHEP-IME $1\times2$ LGAD array}
\end{subfigure}
\hfill
\begin{subfigure}[c]{0.495\textwidth}
\includegraphics[width=\columnwidth, trim=5 40 10 25, clip]{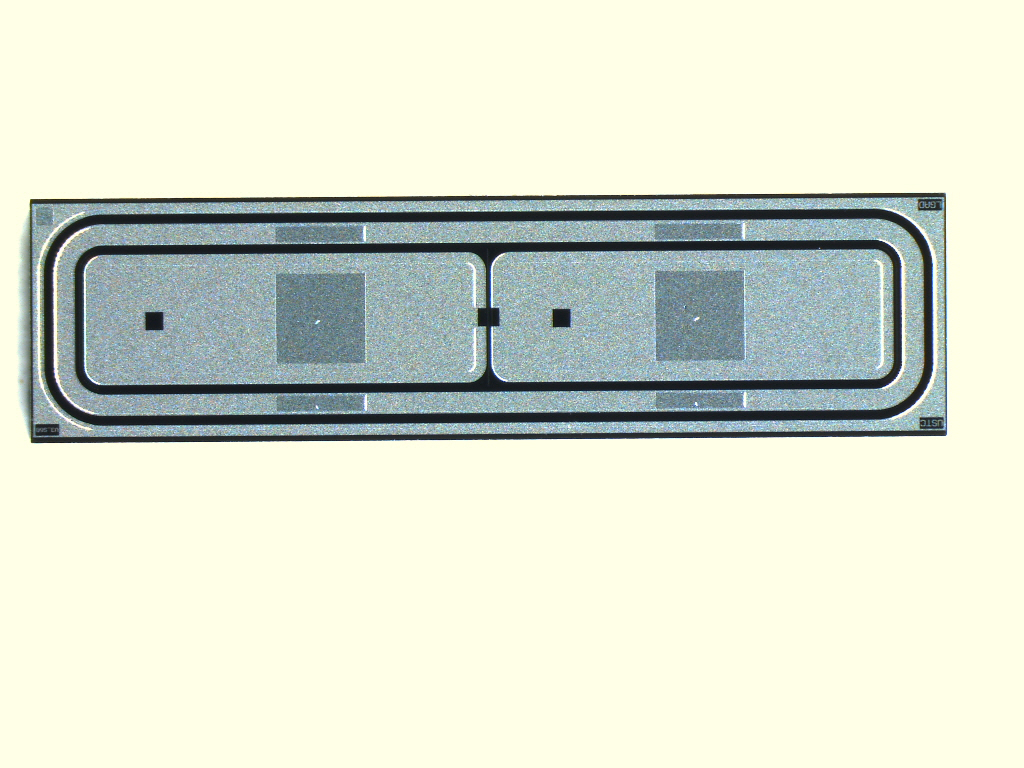}%
\caption{USTC-IME $1\times2$ LGAD array}
\end{subfigure}
\caption{a) Schematics of a Quality Control Test Structure (QCTS). The $1\times2$ LGAD array used in IT is on the far right side. The zoomed in $1\times2$ LGAD structure is shown for: b) IHEP-IME with a single optical window; and c) USTC-IME with three optical windows.}
\label{fig:qcts}
\end{figure}

IT for ATLAS-HGTD is performed on a device called Quality Control Test Structure (QCTS, Figure \ref{fig:qcts}) -- an elongated structure with a size of $21\up{mm}\times1.5\up{mm}$, which is manufactured on an edge of each $15\times15$ sensor on a wafer. The QCTS contains several different structures for control of different process parameters during the production. Among these are also a PIN diode, a single LGAD cell and a group of two LGADs joined along one edge ($1\times2$ LGAD array), which is used in the IT. Each LGAD on the QCTS has a size of $2.1\up{mm} \times 0.8\up{mm}$ which yields the same geometric capacitance as a pixel on the main sensor. The high voltage for sensor biasing is applied between the metallized backplane and the readout electrodes. The top metallization of both LGADs in the $1\times2$ LGAD array has an opening in the interpad region to allow probing with a laser beam. The IHEP-IME design has a single optical window with a size of $300\mum\times200\mum$, while the USTC-IME design has three optical windows with a size of $130\mum\times100\mum$, placed above each LGAD and the interpad region. 

\begin{figure}%
\centering
\begin{subfigure}[c]{0.455\textwidth}
\includegraphics[width=\columnwidth]{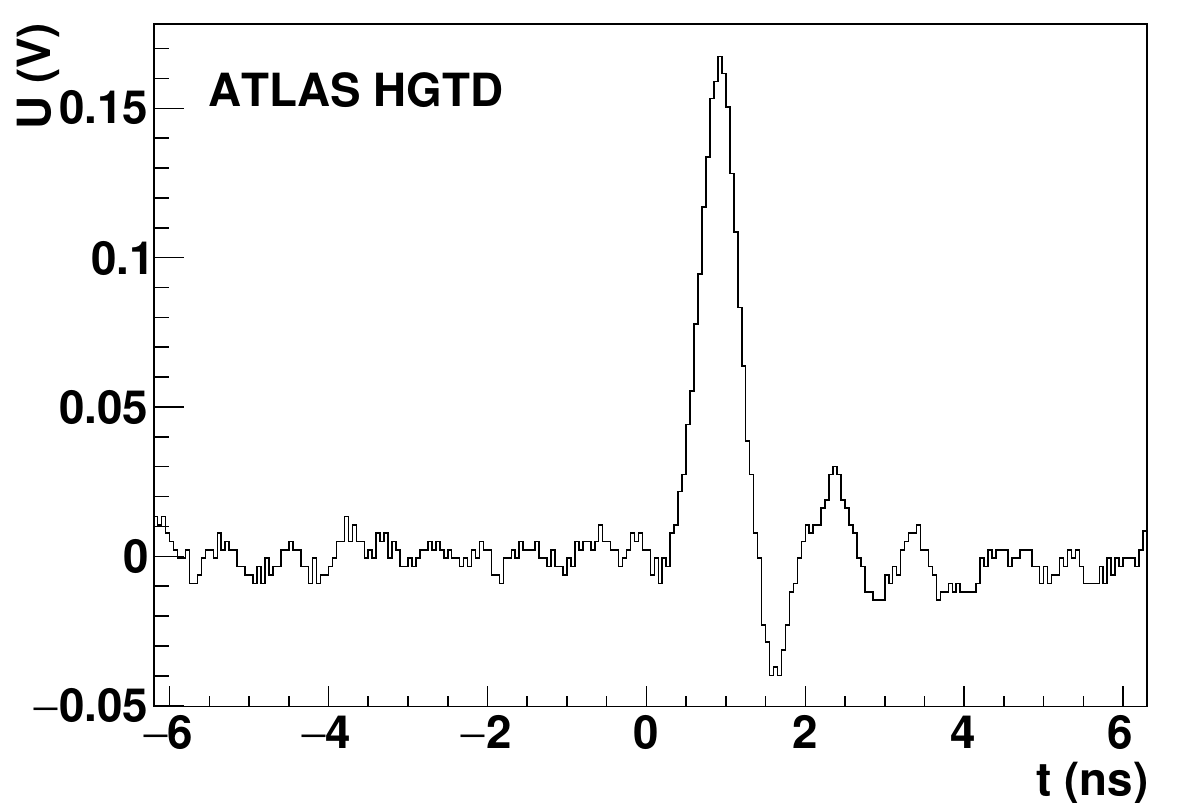}
\caption{}
\label{fig:sr90_pulse}
\end{subfigure}
\begin{subfigure}[c]{0.535\textwidth}
\includegraphics[width=\columnwidth]{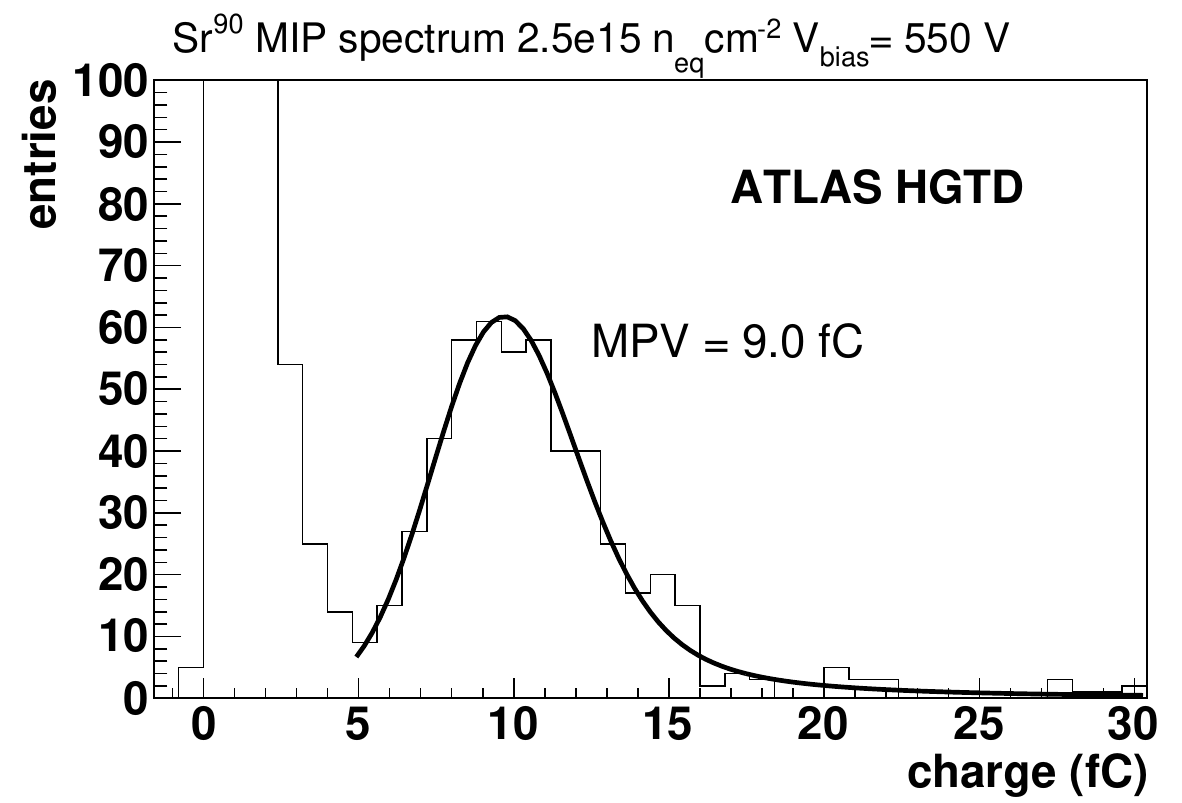}
\caption{}
\label{fig:sr90_spectrum}
\end{subfigure}
\caption{
a) Example LGAD transient signal waveform from the QCTS. In the TCT measurements the charge is extracted by integration over the interval [0, $3\up{ns}$]. In the $\strontium$ measurement the charge is extracted from the waveform maximum in the interval [$-2\up{ns}$, $5\up{ns}$], and is converted to charge using a calibration factor of $80\,\mathrm{fC}/\mathrm{V}$.
b) LGAD charge spectrum from the $\strontium$ measurement. The signal peak is fitted with the Landau-Gaussian function. The peak at $0\up{fC}$ corresponds to triggered events where the particle does not pass the sensitive volume, and is related to noise level (typically $\sigma = 0.7\up{fC}$). Its displacement from 0 is due to the bias in signal sampling.
}
\label{fig:sr90}%
\end{figure}

The IT method has been derived from a method for measurement of the LGAD interpad distance \cite{petja_interpad} and is based on the measurement of charge collection in the LGADs and the interpad region. Tests are done with a Particulars\footnote{Particulars Ltd., Dom\v zale, Slovenia} scanning TCT system at $20\,^\circ\mathrm{C}$ with active temperature stabilization. A QCTS in a dedicated housing is mounted on a 3-axis positioning stage with a submicrometer step resolution. Both pads in the $1\times2$ LGAD array are wire bonded to the same electrical connector and connected to an amplifier. The concept of the measurement is illustrated in Figure \ref{fig:lgad_interpad}. A pulsed beam of infrared light (pulse width $350\up{ps}$, repetition rate $500\up{Hz}$, $\lambda=1064\up{nm}$, penetration depth in Si $\sim1\up{mm}$), focused to a full width at half maximum (FWHM) of $10\mum$, is directed into the interpad optical window from the top side (Top-TCT). Electron-hole pairs are generated along the entire beam path similarly to ionization by a MIP. A beam intensity monitor based on a beam splitter and a photodiode is used for a relative measurement of injected charge and an offline correction of beam intensity fluctuations. Depending on the beam position, the drifting charge carriers are multiplied in the gain layer, or bypass the gain layer through the interpad region. In both cases electrons end their drift on the readout electrode and typical LGAD or PIN responses are generated respectively. The electrical current pulses induced on the readout electrodes by the drifting charge carriers (typical duration $<2\up{ns}$) are amplified using a high bandwidth transimpedance amplifier (Particulars, bandwidth $3\up{GHz}$, gain $53\up{dB}$) and digitized with an oscilloscope (DRS4, bandwidth $700\up{MHz}$). 100 waveforms are averaged at each position to reduce noise. The recorded signals are integrated over the interval of $[0, 3\up{ns}]$ relative to the beginning of the pulse, with the result being proportional to the collected charge (Figure \ref{fig:sr90_pulse}). 

Besides the TCT method, samples have also been characterized with the established C-V technique and in charge collection measurements using a radioactive source $\strontium$. The C-V measurements to determine the LGAD $\vgl$ are carried out at $20\,^\circ\mathrm{C}$ in a probe station using probe needles for contacting. One of the pads in the $1\times2$ LGAD array is probed, while the other pad and the surrounding guard ring are connected to ground and are decoupled from the signal probe. The sensor capacitance as a function of bias voltage is measured at an AC test frequency of $10\up{kHz}$, using the Cp-Rp (parallel capacitance, parallel resistance) model for impedance analysis. 
C-V measurements on irradiated devices are always performed with the sample biased to high voltage at least twice after irradiation. This is because the depletion voltage of the LGAD gain layer, measured during the first biasing, is lower than in subsequent measurements taken immediately (minutes) afterwards. 
The measured value stabilizes after the first biasing cycle.
This effect has been consistently observed on sensors from different manufacturers during the research and development phase and its origin is still under investigation.

Calibrated measurements of collected charge and time resolution with $\strontium$ have been done with a setup described in \cite{3d_timing}, using beta electrons as a substitute for MIPs. The setup uses a reference LGAD with a time resolution of $30\up{ps}$ to determine the time resolution of the investigated sample (DUT). The data acquisition is triggered by a coincidence of signals in the reference LGAD in front of the DUT and a scintillator coupled to a photomultiplier behind the DUT. The DUT signal is not involved in triggering to minimize the trigger bias. Due to geometric trigger acceptance, the particles pass the active part of the DUT in about 5--10$\,\%$ of the triggered events. In a standard measurement 5000 events are recorded of which 400 have a track passing through the DUT. 
The DUT is cooled down to $-30\,^{\circ}\mathrm{C}$ and a single pixel is read out using a discrete transimpedance front end amplifier, described in \cite{SCIPP_board}, and an oscilloscope with a sampling rate of $20\up{GS}/\mathrm{s}$ and an analog bandwidth of $2.5\up{GHz}$.  
In each event the collected charge is measured from the pulse peak of the recorded waveform. The charge spectrum for the entire set of events is fitted above the noise peak with a Landau-Gaussian convolution (Figure \ref{fig:sr90_spectrum}). The most probable value (MPV) of the Landau function is quoted as the collected charge and is converted to absolute charge in fC using a known calibration factor. The time resolution is calculated from the standard deviation of the difference in the time of arrival between the DUT and the reference LGAD, corrected by the time resolution of the reference, using an offline constant fraction discrimination at $20\,\%$ of the pulse maximum to determine the time of arrival.

Samples characterized in this study come from 5 wafers of USTC-IME and 22 wafers of IHEP-IME design, representing first $5\,\%$ of ATLAS-HGTD sensor production (called preproduction). Sample irradiations have been carried out with neutrons at Jo\v{z}ef Stefan Institute's TRIGA reactor \cite{triga1,triga2}. The majority of the samples were irradiated to a fluence of $2.5\times10^{15}\neqcm$, while some were irradiated to lower or higher fluences (up to $3\times10^{15}\neqcm$) to study fluence dependence of the results. The delivered neutron fluence is within $10\,\%$ of the nominal value \cite{sola_reactor}. Samples were annealed for 80 minutes at $60\,^\circ\mathrm{C}$ and otherwise kept in a freezer before measurements. During ATLAS-HGTD production a standard neutron fluence of $2.5\times10^{15}\neqcm$ will be used in all IT.

\begin{figure}%
\centering
\begin{subfigure}[c]{\textwidth}
\includegraphics[width=\columnwidth]{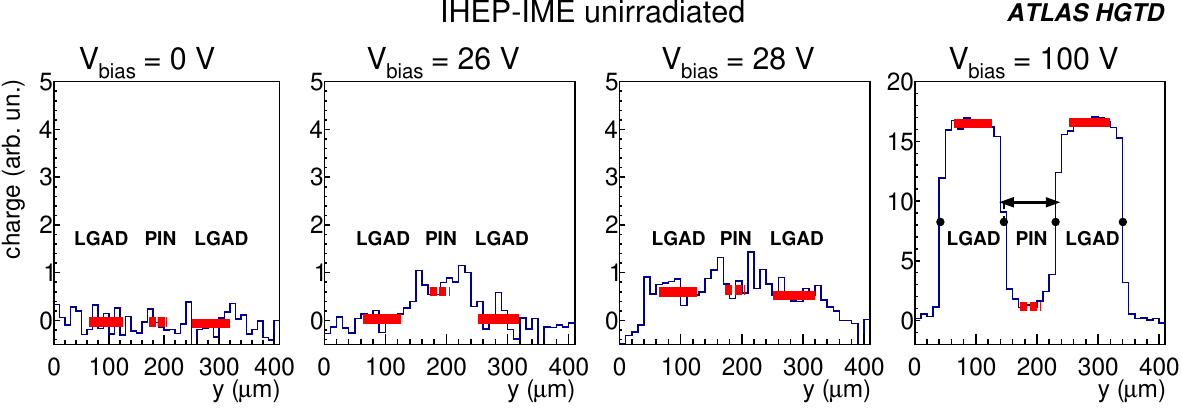}
\caption{}
\label{fig:charge_profile_ihep}
\end{subfigure}
\\
\begin{subfigure}[c]{\textwidth}
\includegraphics[width=\columnwidth]{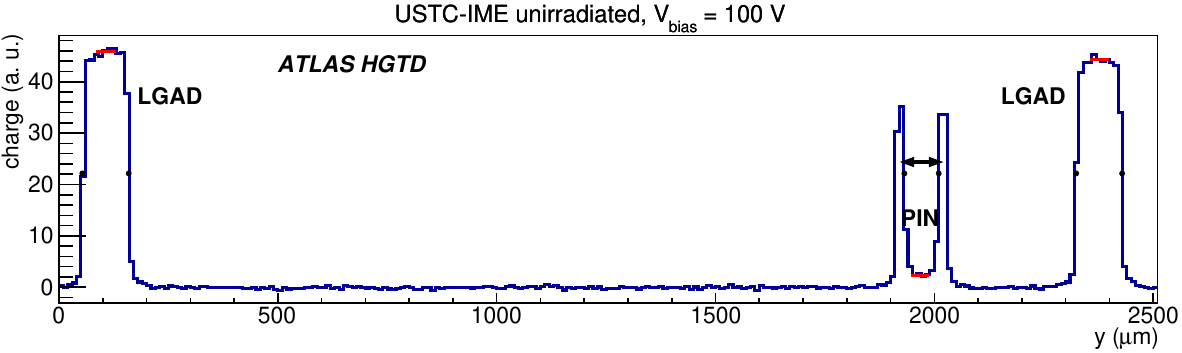}
\caption{}
\label{fig:charge_profile_ustc}
\end{subfigure}
\caption{a) TCT charge collection profiles measured in an IHEP-IME sample at different bias voltages (the $y$-scale at $100\up{V}$ is extended by a factor of 4 for better visibility). The interpad region is centered around $y=190\mum$. The LGAD and PIN signals are fitted with constant functions in the corresponding intervals. At $y<50\mum$ and $y>350\mum$ no signals are generated in the sensor due to beam clipping on surface metallization. b) Charge profile in a USTC-IME sample. Three optical windows are scanned to measure charge in each LGAD and the PIN. Dots in both profiles at $\vbias=100\up{V}$ indicate the extracted rising and falling edges used for defining fit intervals and the arrows indicate the interpad distance.}
\label{fig:charge_profiles}%
\end{figure}


\section{Extraction of LGAD performance parameters from TCT measurements}
\label{sec:parameter_extraction}

TCT measurements are carried out across the LGAD-interpad-LGAD interface at different bias voltages $\vbias$. The beam is scanned over the  $300\mum$ long optical window centered across its width.  Figure \ref{fig:charge_profile_ihep} shows resulting charge collection profiles for different $\vbias$ in an unirradiated IHEP-IME sample with a breakdown voltage of $200\up{V}$. 
At $\vbias=0\up{V}$ no signals are observed. At $V_\mathrm{bias}=\vgl=26\up{V}$, a sizeable signal is detected in the interpad region (PIN diode), which is significantly depleted, while no signals are detected from the LGADs. 
This is because only the $\sim1\mum$ thick LGAD gain layer is depleted, while the bulk is not, hence only a small fraction of generated charge carriers is collected by drift. At $\vbias=28\up{V}$ the LGAD signal reaches the level of the PIN signal. At this point the gain layer, as well as part of the LGAD bulk are depleted and a sizable fraction of the charge generated in the LGAD is collected and multiplied. Finally, at $\vbias=100\up{V}$ the LGAD bulk is fully depleted - the LGADs are in an operating regime and provide highly amplified signals, while the interpad signal remains practically unchanged. The measured width of the charge collection profiles between the first rising edge and the last falling edge is $300\mum$, which agrees with the dimension of the optical window.

A good alignment of the beam with the optical window is essential for precise measurements. The small size of the beam ($10\mum$) compared to the width of the optical window ($100\mum$ or $200\mum$) allows a relatively straightforward alignment with the center of the optical window. The beam position is verified on both ends of the window before the measurement to ensure central alignment. In addition, a data quality control step is executed during analysis to spot potential anomalies in the measurements. The charge profile taken at $\vbias=100\up{V}$ (Figure \ref{fig:charge_profile_ihep}) is examined to determine the beam spot size using the knife-edge technique to verify that the sample is in the beam focus. The procedure also monitors the relative signal sizes from both LGADs, as a smaller signal from one device would indicate the beam leaving the optical window of that device.

The charge collection profiles in Figure \ref{fig:charge_profile_ihep} are fitted with a constant function in the intervals corresponding to the location of each LGAD and the PIN diode. The collected charge and its uncertainty are represented by the fitted value and the error of the fit respectively. The fit intervals are the same for all voltages and are determined from the charge profile at the maximal $\vbias$. In this profile, the positions of rising and falling edges at $50\,\%$ of the maximal profile height are determined by linear interpolation (points marked by four dots in the profile at $\vbias=100\up{V}$ in Figure \ref{fig:charge_profile_ihep}). The outer two points represent the boundaries of the optical window and the inner two points represent the LGAD-PIN interface. 
The intervals between those four points define three fitting regions, corresponding to LGAD-PIN-LGAD respectively. The intervals are reduced for fitting by $20\mum$ on each side to account for edge effects and finite beam width. 

The measurement with USTC-IME samples, shown in Figure \ref{fig:charge_profile_ustc}, is slightly different, because it uses three optical windows instead of one. Since the interpad window is not sufficiently large for beam to access the LGAD region with full gain, the LGAD charge has to be extracted from the two windows in the central LGAD regions.
Fit regions are defined using the same rising/falling edge procedure as described for IHEP-IME devices.
The extracted charge from each window corresponds to LGAD--PIN--LGAD respectively, and does not differ from the outcome of the original method. The measured sizes and positions of the optical windows agree with the physical dimensions of the sample.

\begin{figure}%
\centering
\begin{subfigure}[c]{\textwidth}
\includegraphics[width=0.495\columnwidth, trim=0 220 280 0, clip]{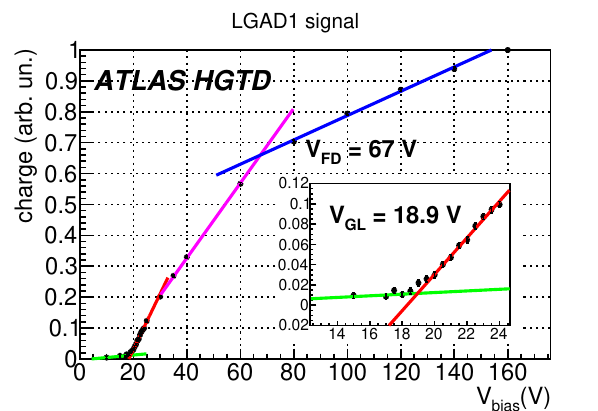}
\includegraphics[width=0.495\columnwidth, trim=280 220 0 0, clip]{plots/tct_analysis_irrad.pdf}
\caption{}
\label{fig:q_vs_vbias_lgads}
\end{subfigure}
\\
\begin{subfigure}[c]{0.495\textwidth}
\includegraphics[width=\columnwidth, trim=0 10 280 210, clip]{plots/tct_analysis_irrad.pdf}
\caption{}
\label{fig:q_vs_vbias_pin}
\end{subfigure}
\begin{subfigure}[c]{0.495\textwidth}
\includegraphics[width=\columnwidth, trim=280 10 0 210, clip]{plots/tct_analysis_irrad.pdf}
\caption{}
\label{fig:q_vs_vbias_gain}
\end{subfigure}
\caption{Example of TCT IT results on a QCTS sample irradiated to $2.5\times10^{15}\neqcm$: a)~Dependence of the LGAD signal size on $\vbias$ in each LGAD pad in the $1\times2$ array. LGAD1 denotes the pad further away, and LGAD2 denotes the pad closer to the single cell LGAD on the QCTS. Intersections of the linear fits are used to determine $\vgl$ (around $20\up{V}$) and LGAD full depletion voltage (around $70\up{V}$). The insets are focused on the region around $\vgl$;  b)~PIN diode signal from the interpad region and extraction of the PIN full depletion voltage from a linear fit intersection; c)~LGAD gain dependence on $\vbias$ as the ratio of the LGAD and PIN signals (markers for LGAD1 and LGAD2 coincide almost completely).}
\label{fig:q_vs_vbias_main}%
\end{figure}

The collected charge in each LGAD and the PIN as a function of $\vbias$ is shown in Figure \ref{fig:q_vs_vbias_main}, where LGAD1 denotes the pad further away from the single LGAD cell and LGAD2 denotes the pad next to the single LGAD cell on the QCTS (Figure \ref{fig:qcts}). In the LGADs the collected charge at $\vbias<\vgl$ is negligible, while above $\vgl$ it increases in two stages: the first stage corresponds to bulk depletion and saturates roughly when the LGAD becomes fully depleted, while the second stage corresponds to an exponential gain increase with bias voltage. 
$\vgl$ is defined as the voltage with the onset of the LGAD signal and is extracted from the intersection point of two linear functions fitted to the data below and above $\vgl$ respectively. 

The selection of the interval for linear fits is a source of a systematic uncertainty on the $\vgl$ value. Below $\vgl$ the fit interval is $[0,15\up{V}]$ and the $\vgl$ uncertainty is minimally affected by the fit boundaries. Above $\vgl$ the fit is done in three iterations in the fit interval of $[V_\mathrm{gl, i-1}+1\up{V}, V_\mathrm{gl, i-1} + 5\up{V}]$, where $V_\mathrm{gl, i-1}$ is the $\vgl$ value obtained from the previous iteration. The initial value $V_\mathrm{gl, 0}$ is picked at average value of $\vgl$ for the given fluence. The $\chi^2$ value of the fit normalized by the number of degrees of freedom ($\chi^2/\mathrm{NDOF}$) is monitored to ensure fit quality. A value of $\chi^2/\mathrm{NDOF}\gg1$ indicates a problematic fit and the fit boundaries are manually adjusted. The typical uncertainty of the $\vgl$ obtained by this procedure is $\pm0.1\up{V}$.

During the bulk depletion the collected charge increases approximately linearly with $\vbias$. The dependence on $\vbias$ is influenced by several factors: the ballistic deficit arising from a short integration time of $3\up{ns}$; the change of the effective bulk space charge concentration with irradiation; and the relocation of the back side electrode of the weighting (Ramo) field to the rear of the sensitive volume after irradiation. 

The collected charge in the PIN diode linearly increases with $\vbias$ up to the point of full depletion of the bulk. Above the full depletion voltage it is roughly constant - the small increase with $\vbias$ is due to the small size of the device and probably comes from stray charge multiplication in the LGAD. 

The bias voltage required for the full depletion of the bulk below the LGAD and the PIN ($\vfd$) is also determined from the intersection of two linear fits in the diagrams in Figure \ref{fig:q_vs_vbias_main}. The fit intervals are selected manually below and above the knee, where data is linear. The estimated uncertainty of the extracted value is $2\up{V}$ after end-of-life fluence. 

The LGAD gain factor $G$ is calculated at different bias voltages $\vbias$ as the ratio of the collected charge in the LGAD and the PIN:
\begin{equation}
    G(\vbias) = \frac{\mathrm{charge}_\mathrm{LGAD}(\vbias)}{\mathrm{charge}_\mathrm{PIN}(\vbias)}.
\end{equation}
The gain at a fixed voltage of $\vbias=100\up{V}$, called $\gfixed$, is taken as a figure of merit in the IT. 

The size of the inactive region between two LGADs (interpad distance) is measured in the charge profile at $\vbias=100\up{V}$. It is defined as the distance between the interpad edges of the LGAD charge profiles at $50\,\%$ of maximal profile height (inner two markers in Figure \ref{fig:charge_profile_ihep} at $\vbias=100\up{V}$).

\section{TCT results and correlation with the other methods}

\begin{figure}%
\centering
\begin{subfigure}[c]{0.49\columnwidth}
\includegraphics[width=\columnwidth, trim=0 0 0 37, clip]{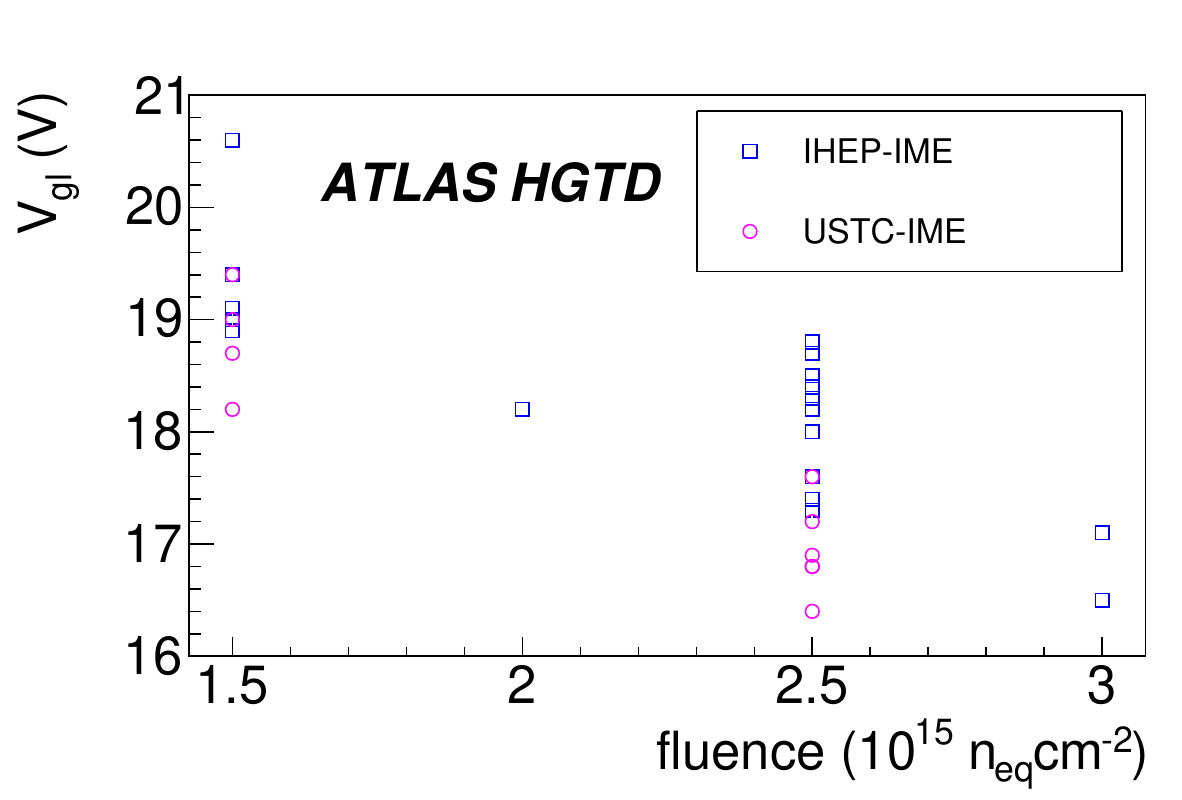}
\caption{}
\label{fig:fluence_vgl}
\end{subfigure}
\begin{subfigure}[c]{0.49\columnwidth}
\includegraphics[width=\columnwidth, trim=0 0 0 36, clip]{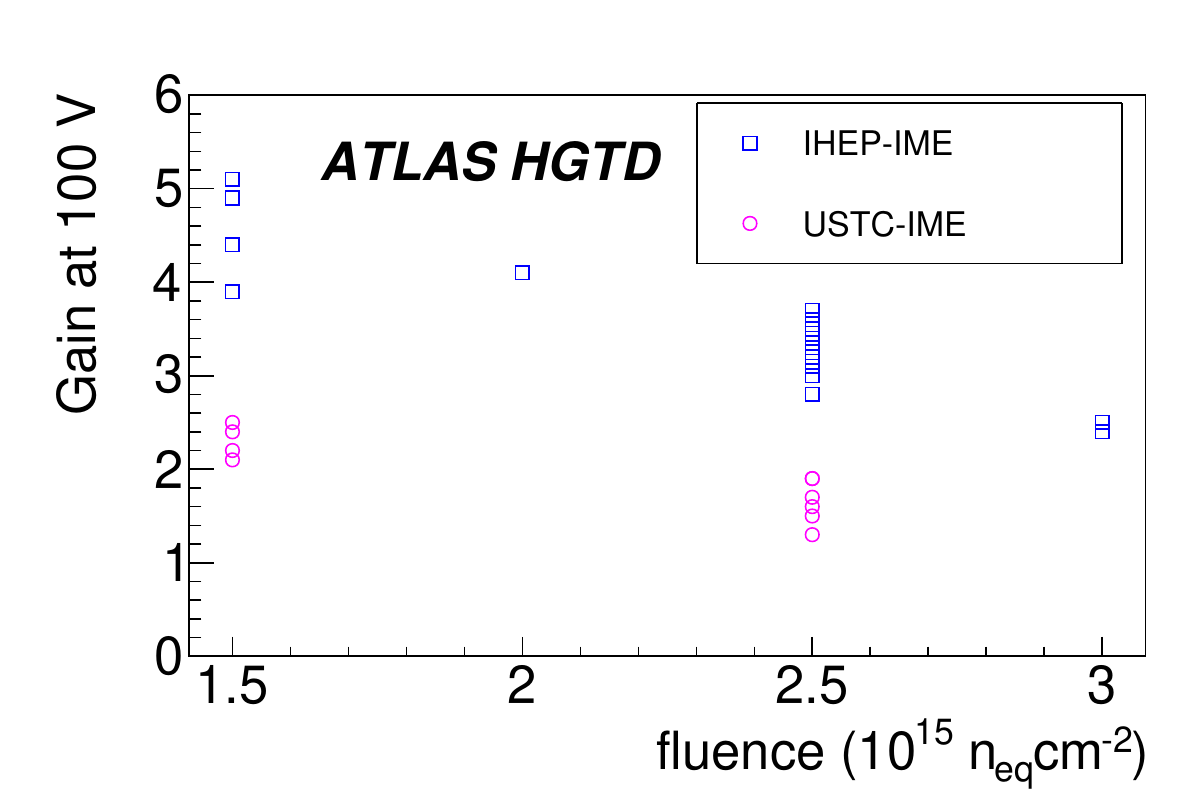}
\caption{}
\label{fig:fluence_gain}
\end{subfigure}
\caption{TCT IT results for a) $\vgl$ and b) $\gfixed$ as a function of fluence.}
\label{fig:tct_fluence}%
\end{figure}

A few QCTS from each of the 27 ATLAS-HGTD preproduction wafers were irradiated with neutrons and tested with the TCT IT, C-V and with $\strontium$ MIPs in order to calibrate the TCT method. The main parameters used in the comparison between the methods are $\vgl$, $\gfixed$ and collected charge from a MIP. Figure \ref{fig:tct_fluence} shows the dependence of $\vgl$ and $\gfixed$ on the fluence measured with TCT. Both parameters distinctly reduce with irradiation, which is expected due to acceptor removal. 
The results within each design vary moderately between samples -- the main causes of the uncertainty and their estimated magnitudes are:
\begin{itemize}
    \item Neutron fluence variation within $\pm10\,\%$, resulting in $\vgl$ peak-to-peak variation of $1.1\up{V}$ after the end-of-life fluence;
    \item Wafer-to-wafer variation: ATLAS-HGTD accepts sensors with the breakdown voltage within $165\up{V}<V_\mathrm{bd}<195\up{V}$ before irradiation, corresponding to a $\vgl$ peak-to-peak variation of $0.5\up{V}$ after the end-of-life fluence;
    \item Manufacturing variations on a wafer - subject to the same initial variation conditions as wafer-to-wafer variations, although the magnitude is typically smaller.
\end{itemize}
The observed higher values of $\vgl$ and, in particular, $\gfixed$ in the IHEP-IME samples suggest a difference in the doping concentration and thickness of the gain layer between the two designs. This does not inherently imply a difference in the radiation hardness of either design, which is addressed in more detail in Section \ref{sec:acceptance_criteria}.

\begin{figure}%
\centering
\begin{subfigure}[c]{0.49\columnwidth}
\includegraphics[width=\columnwidth]{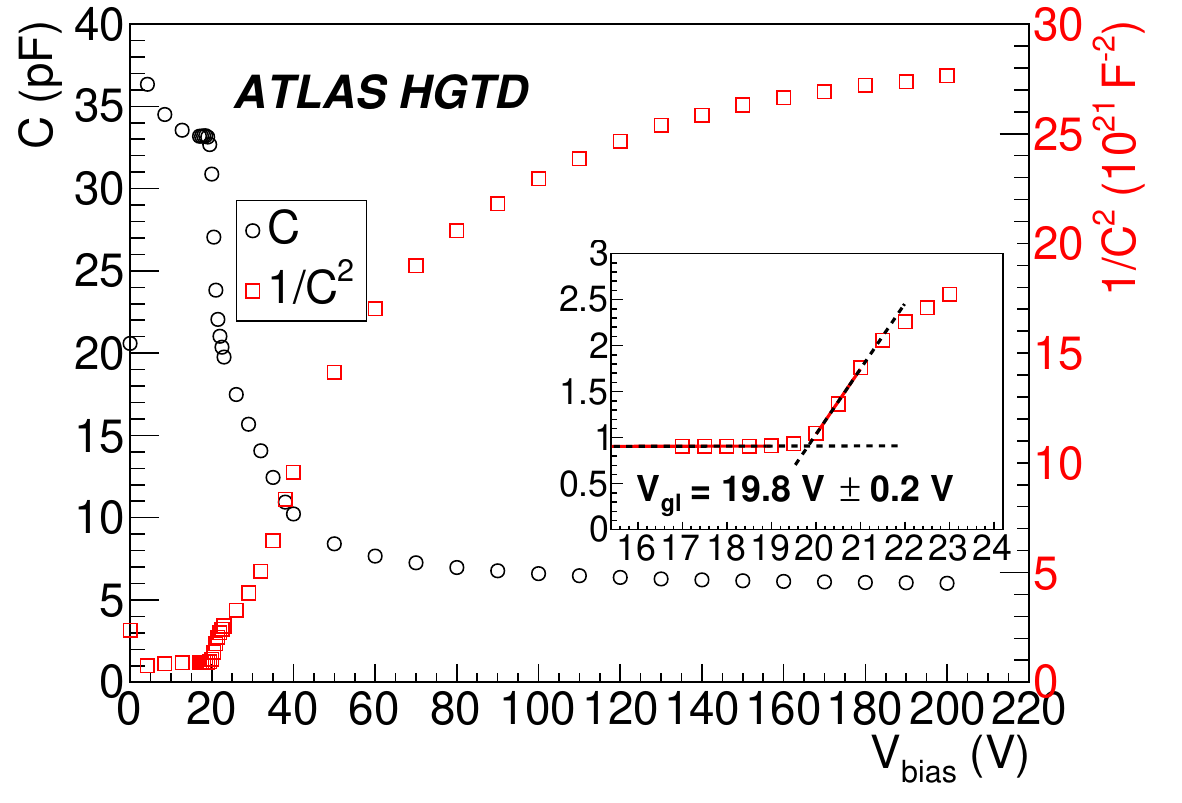}
\caption{}
\label{fig:cv_diagram}
\end{subfigure}
\begin{subfigure}[c]{0.49\columnwidth}
\includegraphics[width=\columnwidth]{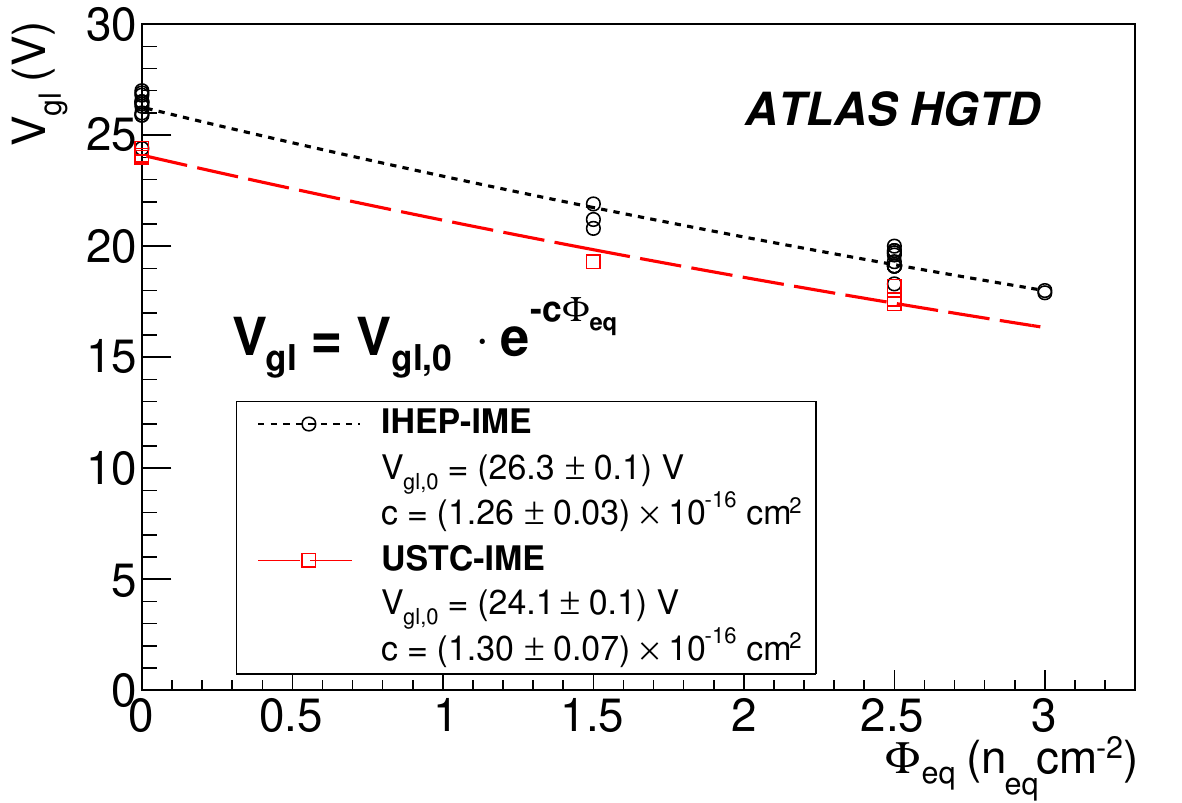}
\caption{}
\label{fig:acceptor_removal}
\end{subfigure}
\begin{subfigure}[c]{0.55\columnwidth}
\includegraphics[width=\columnwidth]{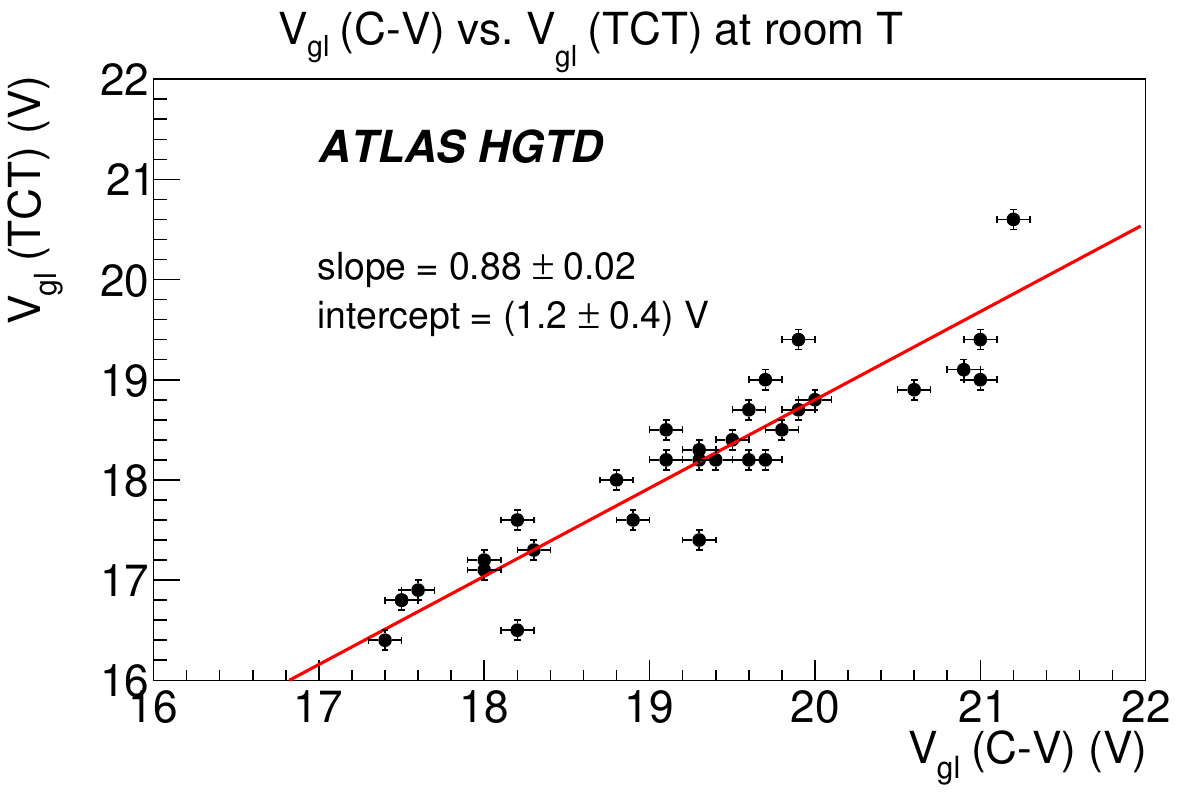}
\caption{}
\label{fig:cv_vgl}
\end{subfigure}
\caption{Summary of C-V measurements: a) Typical LGAD C-V measurement ($\Phi_\mathrm{eq} = 2.5\times10^{15}\neqcm$) showing $C$ and $1/C^2$ dependence on $\vbias$. The inset shows $\vgl$ extraction from the intersection of linear fits. b) $\vgl$ dependence on fluence for IHEP-IME and USTC-IME designs with extraction of the acceptor removal parameter $c$ from an exponential fit of the data. c) Correlation between $\vgl$ values extracted with C-V and TCT methods. Each point corresponds to one neutron irradiated sample. The error bars are the $\vgl$ extraction uncertainties from the respective method.}
\label{fig:cv}%
\end{figure}

The results of the C-V measurements and the correlation with the TCT method are shown in Figure \ref{fig:cv}. 
In C-V, the $\vgl$ is extracted from the dependence of $1/C^2$ on $\vbias$ using the intersection of two linear functions fitted to the data below and above $\vgl$ in manually adjusted fit intervals (Figure \ref{fig:cv_diagram}). 
The dependence of the extracted $\vgl$ on fluence is shown in Figure \ref{fig:acceptor_removal}. The values, which reduce with fluence due to acceptor removal, are fitted with the function:
\begin{equation}
    \vgl = V_\mathrm{gl,0}\times e^{-c\Phi_\mathrm{eq}},
\end{equation}
where $V_\mathrm{gl,0}$ is the $\vgl$ value before irradiation and $c$ is the acceptor removal parameter that describes the radiation hardness of the gain layer. Data from both designs yields the acceptor removal parameter value around $1.3\times10^{-16}\up{cm}^2$. The same value for both designs despite different $V_\mathrm{gl,0}$ can be qualitatively explained by the distinct gain layer doping profile of each design.
The comparison of the $\vgl$ value extracted from individual samples using both the C-V and the TCT method is shown in Figure \ref{fig:cv_vgl}.
The value obtained from TCT is consistently lower than that from C-V, which is due to intrinsic differences between the two methods -- in TCT, $\vgl$ is derived from charge collection, while C-V measures electrical properties that are affected by trapping times and the applied AC frequency, which  are difficult to model. The data points are fitted with a linear function which shows a good correlation between both methods (fit slope value of 0.84). 
This correlation with the established C-V method confirms the sensitivity of the TCT measurement for $\vgl$. Moreover, since the TCT measurement is based on direct charge collection, it is not affected by electrical effects in irradiated samples which are difficult to model in C-V measurements and can lead to different extracted values. Therefore, the resulting $\vgl$ values with TCT are likely more precise.

\begin{figure}
    \centering
    \begin{subfigure}[c]{0.8\columnwidth}
        \includegraphics[width=\columnwidth]{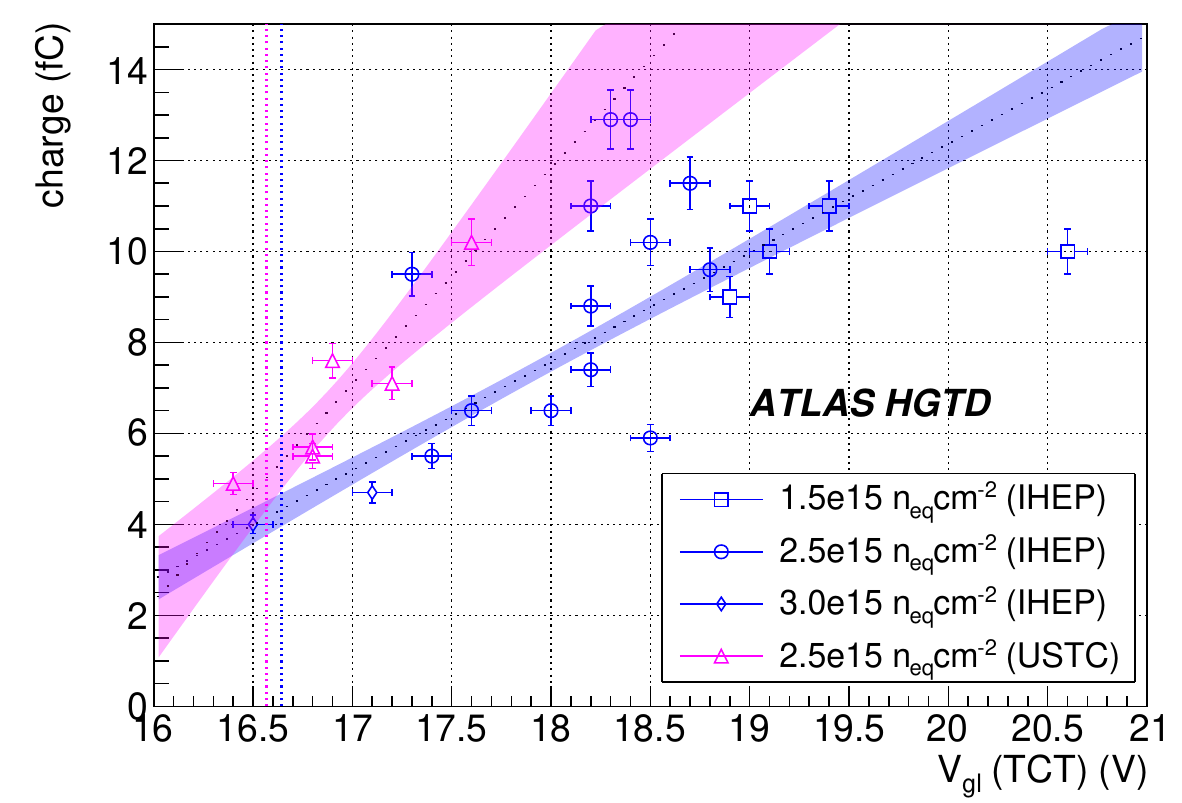}
        \caption{}
        \label{fig:vgl_charge}
    \end{subfigure}
    \begin{subfigure}[c]{0.8\columnwidth}
        \includegraphics[width=\columnwidth]{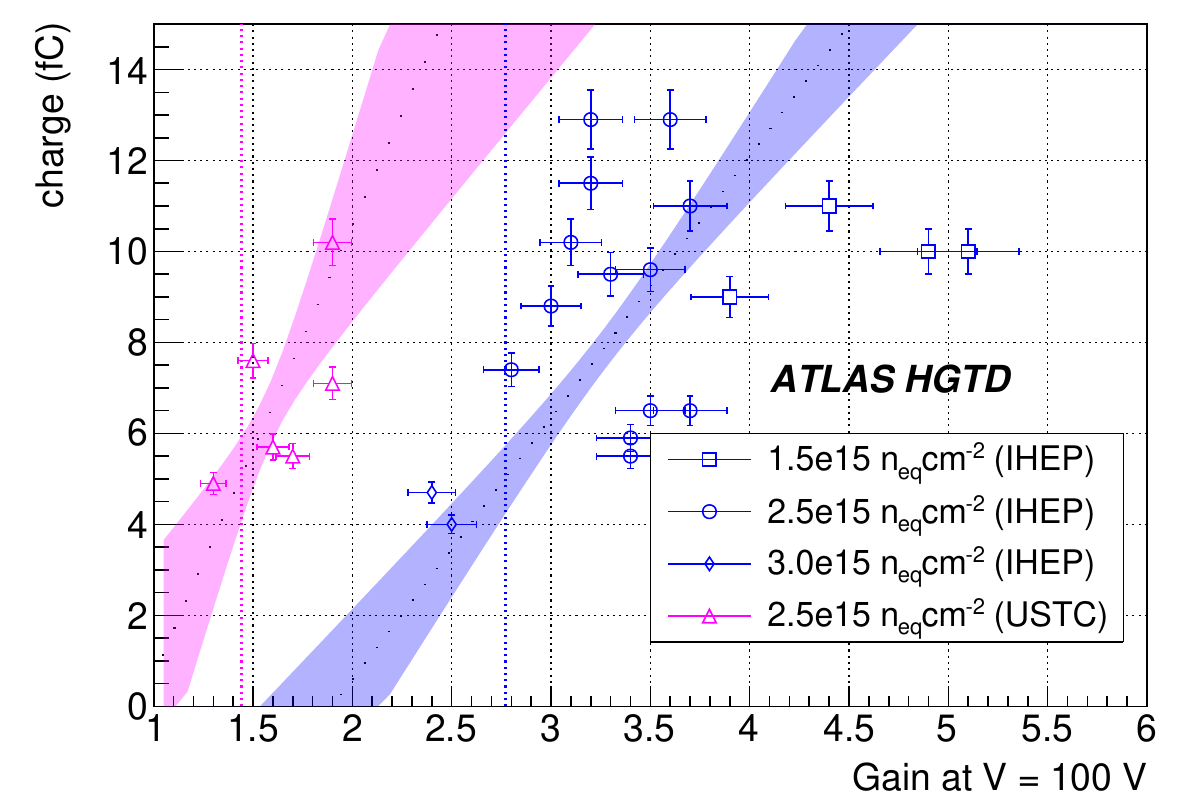}
        \caption{}
        \label{fig:gain_charge}
    \end{subfigure}
    \caption{Correlation between most probable collected charge from $^{90}\mathrm{Sr}$ MIP and a) $\vgl$ and b) $\gfixed$. Each data point represents one IHEP-IME or USTC-IME sample. Data for each design is fitted with a linear fit and uncertainty band represents $1\sigma$ fit parameter variation. The wafer acceptance threshold is based on the point of the fit crossing the charge of $5\up{fC}$, indicated by dashed vertical lines in corresponding colors. }
    \label{fig:correlation_charge}
\end{figure}

The main calibration of the TCT for IT comes from the correlation of $\vgl$ and $\gfixed$ against collected charge with $\strontium$ MIPs. 
The former is similar to a previous study on unirradiated samples that demonstrated a good correlation between $\vgl$ and $\vbias$ required to collect a charge of $20\up{fC}$ \cite{cms-qc}.
The correlation plots are shown in Figure \ref{fig:correlation_charge}. The data points for both designs are grouped by nominal neutron fluence -- for USTC-IME design $\strontium$ measurements were made at the end-of-life fluence of $2.5\times10^{15}\neqcm$. $\strontium$ data for fluences of $2.5\times10^{15}\neqcm$ and above is shown for the operating voltage at the single-event-breakdown limit of $550\up{V}$, while data for the fluence of $1.5\times10^{15}\neqcm$ is shown for an operation voltage about $20\up{V}$ below the sample's breakdown voltage. The whole dataset for each design is fitted with a linear function including an uncertainty band, which represents the $1\sigma$ uncertainty of the fit parameters. 
The correlations show reasonable agreement between the methods, indicating that samples with low $\vgl$ and $\gfixed$ very likely also yield a low amount of charge from a MIP. As already mentioned, $\gfixed$ in the USTC-IME design has a large spread and is less sensitive on charge compared to IHEP-IME. The correlation factor between $\vgl$ and $\gfixed$ is 0.78 for IHEP-IME and 0.93 for USTC-IME samples (not shown on the figure).
A similar comparison was also done between $\vgl/\gfixed$ and $\strontium$ time resolution. The measured time resolution is in the range of 35--$50\up{ps}$ for all irradiated wafers, but the precision is low due to low statistics and the correlation with the TCT measurements is not significant enough to be used for the wafer acceptance decision. 

\section{Discussion of acceptance criteria} 
\label{sec:acceptance_criteria}
The presented correlation plots between the TCT and $\strontium$ measurements are the basis for the definition of the ATLAS-HGTD wafer acceptance criteria. A minimal required $\strontium$ charge of $5\up{fC}$ is defined as the acceptance threshold. This value is $20\,\%$ larger than the ATLAS-HGTD specification of $4\up{fC}$, to account for two differences in collected charged between $\strontium$ beta electrons and MIPs:
\begin{itemize}
    \item $\strontium$ beta electrons deposit about $10\,\%$ more energy in silicon than MIPs, as determined by GEANT4 simulations;
    \item  LGAD gain decreases with increasing deposited charge density due to electric field screening in the gain layer \cite{curras,kramberger_screening}. Since beta electrons undergo more multiple scattering than MIPs, their charge cloud is broader, resulting in a smaller reduction of gain. This effect is conservatively estimated to contribute an additional $10\,\%$ difference.
\end{itemize}
This threshold is mapped into corresponding $\vgl$ and $\gfixed$ thresholds for TCT measurements. If the TCT measurement yields parameters above these thresholds, the wafer is accepted. Otherwise, additional TCT and $\strontium$ measurements are carried out on up to three structures from that wafer and the wafer is accepted or rejected based on the majority of the test outcomes.

Preproduction data indicates acceptance threshold values of $\vgl\geq16.9\up{V}$ and $\gfixed\geq2.7$ for IHEP-IME and $\vgl\geq16.6\up{V}$ and $\gfixed\geq1.4$ for USTC-IME at a charge threshold of $5\up{fC}$. 
 
Figure \ref{fig:correlation_charge} indicates that the required precision of the $\vgl$ measurement around the acceptance threshold is on the level of $\pm0.1\up{V}$. One particular source of systematic uncertainty is the selection of fit intervals for the $\vgl$ extraction (Figure \ref{fig:q_vs_vbias_main}). The fit quality will be monitored in each measurements and the fit intervals will be manually adjusted in case of obvious deviations.

The leakage current and the interpad distance in the samples are also monitored during the IT. Their acceptable range of values is wide enough that most samples are expected to pass the specifications, however automated checks are set up to spot significant anomalies.

Some parameters are not monitored or included in the acceptance decision. 
The time resolution in $\strontium$ measurements was demonstrated to be below $50\up{ps}$ for charge above $5\up{fC}$ in all preproduction samples. Since the main factor in the time resolution is the $S/N$ ratio, the same timing performance is expected for all samples that achieve the $5\up{fC}$ charge requirement when using the same readout electronics.
However, due to the low statistics/precision of the measurements the results are not used for matching with the TCT results. The TCT measurement of the sensor full depletion voltage in the PIN/LGAD devices varies significantly between samples and is not a parameter which will be included in the IT decision. 

The granularity of the quality control irradiation tests of one to three samples per wafer is relatively low on its own to guarantee the sufficient radiation hardness of all main sensors on the wafer. However, the main sensors undergo additional quality control tests during manufacturing (without irradiation) to ensure device uniformity. In these tests each column of $15\times1$ pixels undergoes an I-V measurement which determines the breakdown voltage and by extension the $\vgl$. The sensor specifications allow relatively small variations of these  parameters, which ensures uniform evolution of the pixel performance with irradiation.

When applying the acceptance criteria to the ATLAS-HGTD preproduction batch, all 22 IHEP-IME and 4 out of 5 USTC-IME wafers passed the test outright. One of the USTC-IME devices did not pass the collected charge test ($4.9\up{fC}$) and another device from the same wafer was tested, according to the protocol explained above. The second device passed the test and the wafer was accepted.


\section{Conclusion}

A fast and comprehensive Irradiation Test (IT) method is required in the ATLAS-HGTD sensor production to monitor the radiation hardness on wafer level, where a collected MIP charge of $4\up{fC}$ and a time resolution of $50\up{ps}$ are required. We introduce a TCT based charge collection measurement in the interpad region of an irradiated $1\times2$ LGAD array to extract sensor performance parameters: $\vgl$, gain dependence on bias voltage $G(\vbias)$, interpad distance and leakage current. The results of TCT measurements have been compared with C-V measurements and correlated with charge collection measurements with $\strontium$ MIPs, and good agreement has been observed between different methods. Mapping of TCT results to collected charge measured with $\strontium$ has been used to define the acceptance criteria for wafer radiation hardness. The main figures of merit are $\vgl$ and $\gfixed$ after the end-of-life neutron fluence of $2.5\times10^{15}\neqcm$. Their cutoff values have been derived for each of the two sensor designs based on IT results in sensor preproduction. A multi-stage production-wafer acceptance procedure has been defined, ensuring adequate radiation hardness of the delivered ATLAS-HGTD sensors.


\section*{Acknowledgment}

We gratefully acknowledge the financial support from the Slovenian Research and Innovation Agency (ARIS P1-0135, ARIS Z1-50011, ARIS J7-4419), Slovenia;
the National Natural Science Foundation of China (No. 11961141014) and the Ministry of Science and Technology of China (No. 2023YFA1605901), China;
and FAPESP (2020/04867-2, 2022/14150-3, 2023/18486-9 and 2023/18484-6), MCTI/CNPq (INCT CERN Brasil 406672/2022-9) and CAPES - Code 001, Brazil.
We acknowledge the support from BMFTR, Germany; NWO, Netherlands; and FCT, Portugal.

\end{document}